# Power Allocation for Fading Channels with Peak-to-Average Power Constraints

Khoa D. Nguyen, Albert Guillén i Fàbregas and Lars K. Rasmussen

**Abstract**

Power allocation with peak-to-average power ratio constraints is investigated for transmission over Nakagami-$m$ fading channels with arbitrary input distributions. In the case of delay-limited block-fading channels, we find the solution to the minimum outage power allocation scheme with peak-to-average power constraints and arbitrary input distributions, and show that the signal-to-noise ratio exponent for any finite peak-to-average power ratio is the same as that of the peak-power limited problem, resulting in an error floor. In the case of the ergodic fully-interleaved channel, we find the power allocation rule that yields the maximal information rate for an arbitrary input distribution and show that capacities with peak-to-average power ratio constraints, even for small ratios, are very close to capacities without peak-power restrictions.

K. D. Nguyen and L. K. Rasmussen are with the Institute for Telecommunications Research, University of South Australia, Mawson Lakes Boulevard, Mawson Lakes 5095, South Australia, Australia, e-mail: dangkhoa.nguyen@postgrads.unisa.edu.au, lars.rasmussen@unisa.edu.au.

A. Guillén i Fàbregas is with the Department of Engineering, University of Cambridge, Trumpington Street, Cambridge CB2 1PZ, UK, e-mail: guillen@ieee.org.

This work was submitted in part at 2008 IEEE International Symposium on Information Theory, Toronto, Canada, July 2008.

This work has been supported by the Australian Research Council under ARC grants RN0459498, DP0558861 and DP0881160.

# I. INTRODUCTION

The ultimate goal of a wireless communication system is to reliably transport high data rates over channels with time-varying transfer characteristics; commonly termed fading channels [1], [2]. When channel state information (CSI), namely knowledge of the channel realizations, is not readily available to the transmitter, error control coding and automatic-repeat-request (ARQ) techniques have been used extensively to compensate for the fading characteristics of the channels [3]. In some systems, CSI can be made available at the transmitter, either via a dedicated feedback link [4], or using channel reciprocity in systems employing time-division duplex (TDD) [5]. In this case, power and rate can be adapted according to the channel realization to further improve the rate/reliability performance of the system. Different adaptation techniques can be employed depending on the system requirements and the nature of the wireless channel [2]. In this paper, we consider power allocation techniques that minimize the word error rate over slowly-varying fading channels, and maximize the ergodic capacity over fast fading channels [2].

Firstly, we consider systems where codewords are transmitted over channels with $B$ degrees of freedom, where $B$ is finite. Examples for such scenarios are transmission over slowly-varying channel, or transmission using orthogonal division multiplexing (OFDM) techniques over frequency selective channels. The channel is conveniently modeled as a block-fading channel [6], [7], where each codeword is transmitted over $B$ corresponding flat fading blocks. In this case, the maximal achievable rate is a random variable, dependent on the channel realization. For most fading statistics, the channel capacity is zero since there is a non-zero probability that any positive rate is not supported by the channel. A relevant performance measure in this case is the information outage probability [7], which is the probability that communication at a target rate $R$ is not supported by the channel. The outage probability is also a lower bound of the word error probability for communicating with rate $R$ over the channel [8]. In this case, power allocation techniques aim at minimizing the outage probability given a rate $R$. The optimal power allocation problem has been investigated in [8] for channels with Gaussian inputs, and in [9], [10], [11] for channels with arbitrary input constellations. The works in [8], [10] consider systems with peak (per-codeword) power constraints and average power constraints, and show that systems with average power constraints perform significantly better than systems with peak power constraints. However, systems with average power constraints employ very large (possibly infinite) peak power, which is not feasible in practice. To this end, the optimal power allocation strategy for Gaussian input channels with both peak and average power constraints is also derived in [8].

For transmission over a fast-varying fading channel, the fading statistics are revealed within each codeword, and the channel is ergodic, i.e. it has infinite degrees of freedom ($B \to \infty$). In this case, adaptive techniques aim at maximizing the ergodic channel capacity, which is the maximum data rate that can be transmitted over the channel with vanishing error probability [12]. Optimal power allocation schemes, such as water-filling for channels with Gaussian inputs [12], [8] and mercury/water-filling for channels with an arbitrary input [9], have been developed for systems with average power constraints. The work in [13] derives the optimal power allocation strategy for Gaussian input channels with both peak and average power constraint, which results in a variation to the classical water-filling algorithm [12].

In this paper, we consider power allocation strategies for arbitrary input channels with peak-to-average power ratio (PAPR) constraints. We derive the optimal power allocation scheme that minimizes outage probability for transmission with arbitrary inputs over a block-fading channel. The optimal power allocation strategy that maximizes the ergodic capacity for arbitrary input channels is also derived. In both cases, the optimal power allocation strategies rely on the first derivative of the input-output mutual information, which may be computationally prohibitive for implementation in specific low-cost systems. We therefore study a suboptimal power allocation scheme, which significantly reduces the computational and storage requirements, while incurring minimal performance loss compare to the optimal scheme.

The remainder of the paper is organized as follows. Sections II and III describe the system model and the information theoretic framework of the work. Section IV discusses power allocation algorithms for minimizing the outage probability of delay-limited block-fading channels, while algorithms for maximizing the ergodic capacity is given in Section V. Concluding remarks are given in Section VI.

## II. System Model

Consider transmission over a channel consisting of $B$ blocks of $L$ channel uses, in which, block $b, b = 1, \ldots, B$, undergoes an independent fading gain $h_b$, corresponding to a power fading gain $\gamma_b \triangleq |h_b|^2$. Assume that $\boldsymbol{h} = (h_1, \ldots, h_B)$ and $\boldsymbol{\gamma} = (\gamma_1, \ldots, \gamma_B)$ are available at the receiver and the transmitter, respectively. Suppose the transmit power is allocated following the rule $\boldsymbol{p}(\boldsymbol{\gamma}) = (p_1(\boldsymbol{\gamma}), \ldots, p_B(\boldsymbol{\gamma}))$. Then the corresponding complex base-band equivalent is given by

$$\boldsymbol{y}_b = \sqrt{p_b(\boldsymbol{\gamma})} h_b \boldsymbol{x}_b + \boldsymbol{z}_b, \quad b = 1, \ldots, B, \tag{1}$$

where $\boldsymbol{y}_b \in \mathbb{C}^L, \boldsymbol{x}_b \in \mathcal{X}^L$, with $\mathcal{X} \subset \mathbb{C}$ being the signal constellation set, are the received and transmitted signals in block $b$, respectively, and $\boldsymbol{z}_b \in \mathbb{C}^L$ is the additive white Gaussian noise (AWGN) vector with independently identically distributed circularly symmetric Gaussian entries $\sim \mathcal{N}_{\mathbb{C}}(0, 1)$. Assume that the signal constellation $\mathcal{X}$ of size $2^M$ satisfies $\sum_{x \in \mathcal{X}} |x|^2 = 2^M$, then the instantaneous received signal-to-noise ratio (SNR) at block $b$ is given by $p_b(\boldsymbol{\gamma})\gamma_b$. We consider systems with the following power constraints:

$$\text{Peak power}: \ \langle \boldsymbol{p}(\boldsymbol{\gamma}) \rangle \triangleq \frac{1}{B} \sum_{b=1}^{B} p_b(\boldsymbol{\gamma}) \leq P_{\text{peak}},$$

$$\text{Average power}: \mathbb{E}\left[\langle \boldsymbol{p}(\boldsymbol{\gamma}) \rangle\right] \leq P_{\text{av}}.$$

For the fully-interleaved ergodic case, the channel model can be obtained from (1) by letting $B \to \infty$ and $L = 1$. Due to ergodicity, power allocation for block $b$ is only dependent on $\gamma_b$. For simplicity of notation, denote $p(\gamma)$ as the transmit power corresponding to the power fading gain $\gamma$. The following power constraints are considered:

$$\text{Peak power}: p(\gamma) \le P_{\text{peak}},$$
$$\text{Average power}: \mathbb{E}_\gamma\left[p(\gamma)\right] \le P_{\text{av}}.$$

Power allocation schemes for systems with peak power constraints and average power constraints have been studied in [8], [10] for the delay-limited channel and in [14], [13] for the ergodic channel[1]. Power allocation with average power constraints offers significant performance advantage but requires large peak powers [10], [8], which may prohibit application in practical systems. In this work, we study the performance of systems with peak power constraints in addition to average power constraints [8], [13]. In particular, we consider systems with a constrained *peak-to-average power ratio* $\text{PAPR} \triangleq \frac{P_{\text{peak}}}{P_{\text{av}}} \ge 1$.

We consider block-fading channels where the fading gain $h_b$ has Nakagami-$m$ distributed magnitude and uniformly distributed phase[2]. The probability density function (pdf) of $|h_b|$ of the fading gain is

$$f_{|h_b|}(\xi) = \frac{2m^m\xi^{2m-1}}{\Gamma(m)}e^{-m\xi^2}, \quad b = 1,\ldots,B,$$

where $\Gamma(a)$ is the Gamma function, $\Gamma(a) = \int_0^\infty t^{a-1}e^{-t}dt$. The pdf of the power fading gain is then given by

$$f_\gamma(\gamma) = \begin{cases} \frac{m^m\gamma^{m-1}}{\Gamma(m)}e^{-m\gamma}, & \gamma \ge 0 \\ 0, & \text{otherwise.} \end{cases} \tag{2}$$

The Nakagami-$m$ distribution represents a large class of practical fading statistics. In particular, we can recover the Rayleigh fading by setting $m = 1$ and approximate the Ricean fading with parameter $K$ by setting $m = \frac{(K+1)^2}{2K+1}$ [1].

[1] In the literature, peak power constraints have also been referred to as short-term power constraints, and average power constraints as long-term power constraints.

[2] We assume that the phase will be perfectly compensated due to the perfect CSI at the receiver.

## III. Outage Probability and Ergodic Capacity

Let $I_{\mathcal{X}}(\rho)$ be the input-output mutual information of an AWGN channel with input constellation $\mathcal{X}$ and received SNR $\rho$. Given a channel realization $\boldsymbol{\gamma}$ and a power allocation scheme $\boldsymbol{p}(\boldsymbol{\gamma})$ satisfying the power constraint $P$, the instantaneous input-output mutual information of the delay-limited block-fading channel given in (1) is

$$I_B(\boldsymbol{p}(\boldsymbol{\gamma}),\boldsymbol{\gamma}) = \frac{1}{B}\sum_{b=1}^{B} I_{\mathcal{X}}(p_b\gamma_b). \tag{3}$$

For a fixed transmission rate $R$, communication is in outage when $I_B(\boldsymbol{p}(\boldsymbol{\gamma}),\boldsymbol{\gamma}) < R$. The outage probability, which is a lower bound to the word error probability, is given by

$$P_{\text{out}}(\boldsymbol{p}(\boldsymbol{\gamma}),P,R) \triangleq \Pr(I_B(\boldsymbol{p}(\boldsymbol{\gamma}),\boldsymbol{\gamma}) < R). \tag{4}$$

Besides, the capacity of an ergodic fading channel with input constellation $\mathcal{X}$ and power allocation rule $p(\gamma)$ is given by

$$C \triangleq \mathbb{E}_{\gamma}\left[I_{\mathcal{X}}(p(\gamma)\gamma)\right]. \tag{5}$$

The mutual information $I_{\mathcal{X}}(\rho)$ in (3) and (5) is defined as follows. With Gaussian inputs, we have that $I_{\mathcal{X}_G}(\rho) = \log_2(1+\rho)$, while for coded modulation over uniformly-distributed fixed discrete signal constellations[3], we have that

$$I_{\mathcal{X}}(\rho) = M - \frac{1}{2^M}\sum_{x\in\mathcal{X}}\mathbb{E}_Z\left[\log_2\left(\sum_{x'\in\mathcal{X}} e^{-|\sqrt{\rho}(x-x')+Z|^2+|Z|^2}\right)\right],$$

where $Z \sim \mathcal{N}_{\mathbb{C}}(0,1)$. We also consider systems with bit-interleaved coded modulation (BICM) using the classical non-iterative BICM decoder proposed by Zehavi in [15]. The mutual information for a given labelling rule can be expressed as [16]

$$I_{\mathcal{X}}^{\text{BICM}}(\rho) = M - \frac{1}{2^M}\sum_{c=0}^{1}\sum_{j=1}^{M}\sum_{x\in\mathcal{X}_c^j}\mathbb{E}_Z\left[\log_2\frac{\sum_{x'\in\mathcal{X}} e^{-|\sqrt{\rho}(x-x')+Z^2|}}{\sum_{x'\in\mathcal{X}_c^j} e^{-|\sqrt{\rho}(x-x')+Z^2|}}\right], \tag{6}$$

[3] Although the restriction to uniform constellations is very relevant in practice and mathematically convenient, the main results of this paper only depend on the underlying probability distribution on $\mathcal{X}$ through the mutual information and its first derivative. Therefore, the generalization is straightforward.

where the sets $\mathcal{X}_c^j$ contain all signal points where the $j^{\text{th}}$ position in the corresponding binary signal-point labelling is $c$.

In deriving optimal power allocation schemes, a useful measure is the first derivative of the mutual information $I_{\mathcal{X}}(\rho)$ with respect to the SNR [9], [10]. From [17] we have that,

$$\frac{d}{d\rho} I_{\mathcal{X}}(\rho) = \frac{1}{\log 2} \mathrm{MMSE}_{\mathcal{X}}(\rho),$$

where $\mathrm{MMSE}_{\mathcal{X}}(\rho)$ is the minimum mean-square error (MMSE) in estimating an input symbol in $\mathcal{X}$ transmitted over an AWGN channel with SNR $\rho$. For Gaussian inputs, $\mathrm{MMSE}_{\mathcal{X}_G}(\rho) = \frac{1}{1+\rho}$, while for a general constellation $\mathcal{X}$, we have that [9]

$$\mathrm{MMSE}_{\mathcal{X}}(\rho) = \frac{1}{2^M} \sum_{x \in \mathcal{X}} |x|^2 - \frac{1}{\pi} \int_{\mathbb{C}} \frac{\left| \sum_{x \in \mathcal{X}} x e^{-|y-\sqrt{\rho}x|^2} \right|^2}{\sum_{x \in \mathcal{X}} e^{-|y-\sqrt{\rho}x|^2}} dy.$$

For systems with BICM, the first derivative of the mutual information with respect to SNR is given by [18][4]

$$\mathrm{MMSE}_{\mathcal{X}}^{\mathrm{BICM}}(\rho) \triangleq \frac{d}{d\rho} I_{\mathcal{X}}^{\mathrm{BICM}}(\rho) = \sum_{j=1}^{M} \frac{1}{2} \sum_{c=0}^{1} \left( \mathrm{MMSE}_{\mathcal{X}}(\rho) - \mathrm{MMSE}_{\mathcal{X}_c^j}(\rho) \right).$$

In the remainder of the paper, we perform analysis for the coded modulation case. Results for the BICM case can be obtained by simply replacing $I_{\mathcal{X}}(\rho), \mathrm{MMSE}_{\mathcal{X}}(\rho)$ by $I_{\mathcal{X}}^{\mathrm{BICM}}(\rho)$ and $\mathrm{MMSE}_{\mathcal{X}}^{\mathrm{BICM}}(\rho)$, respectively.

## IV. Outage Probability Minimization

### A. Peak and Average Power Constraints

In this section, we review known results on peak-power and average-power constrained systems, respectively, over delay-limited channels relevant to our main results. A detailed treatment of optimal and suboptimal power allocation schemes for systems with peak-power and average-power constraints, respectively, over delay-limited block-fading channels is given in [10].

[4]With some abuse of notation, we use $\mathrm{MMSE}_{\mathcal{X}}^{\mathrm{BICM}}(\rho)$ to denote the first derivative with respect to $\rho$ of the mutual information. However, $\mathrm{MMSE}_{\mathcal{X}}^{\mathrm{BICM}}(\rho)$ is not the minimum mean-square error in estimating the channel input from its output, since the noise is not Gaussian due to the demodulation process.

*1) Peak Power Constraint:* For systems with peak power constraint $P_{\rm peak}$, the optimal power allocation scheme is the solution of the following problem [8]

$$\begin{cases} \text{Minimize} & P_{\rm out}\left(\boldsymbol{p}(\boldsymbol{\gamma}), P_{\rm peak}, R\right) \\ \text{Subject to} & \langle \boldsymbol{p}(\boldsymbol{\gamma})\rangle \le P_{\rm peak} \\ & p_b \ge 0, \quad b = 1,\ldots,B \end{cases} \tag{7}$$

The solution is given by [9], [10]

$$p_b^{\rm peak}(\boldsymbol{\gamma}) = \frac{1}{\gamma_b}\mathrm{MMSE}_{\mathcal{X}}^{-1}\left(\min\left\{\mathrm{MMSE}_{\mathcal{X}}(0), \frac{\eta}{\gamma_b}\right\}\right), \tag{8}$$

for $b = 1,\ldots,B$ where $\eta$ is chosen such that the peak power constraint is met with equality. As shown in [10], an alternative optimal power allocation rule for peak power constraint is given by

$$\boldsymbol{p}^{\rm peak}(\boldsymbol{\gamma}) = \begin{cases} \boldsymbol{\wp}(\boldsymbol{\gamma}), & \text{if } \langle \boldsymbol{\wp}(\boldsymbol{\gamma})\rangle \le P_{\rm peak} \\ \mathbf{0}, & \text{otherwise,} \end{cases} \tag{9}$$

where $\boldsymbol{\wp}(\boldsymbol{\gamma})$ is the solution of the problem

$$\begin{cases} \text{Minimize} & \langle \boldsymbol{\wp}(\boldsymbol{\gamma})\rangle \\ \text{Subject to} & I_B(\boldsymbol{\wp}(\boldsymbol{\gamma}), \boldsymbol{\gamma}) \ge R \\ & \wp_b \ge 0, \quad b = 1,\ldots,B. \end{cases} \tag{10}$$

From [10], $\boldsymbol{\wp}(\boldsymbol{\gamma})$ is given by

$$\wp_b(\boldsymbol{\gamma}) = \frac{1}{\gamma_b}\mathrm{MMSE}_{\mathcal{X}}^{-1}\left(\min\left\{\mathrm{MMSE}_{\mathcal{X}}(0), \frac{1}{\eta\gamma_b}\right\}\right), \quad b = 1,\ldots,B \tag{11}$$

where $\eta$ is now chosen such that the rate constraint is met,

$$\frac{1}{B}\sum_{b=1}^{B} I_{\mathcal{X}}\left(\mathrm{MMSE}_{\mathcal{X}}^{-1}\left(\min\left\{\mathrm{MMSE}_{\mathcal{X}}(0), \frac{1}{\eta\gamma_b}\right\}\right)\right) = R.$$

The power allocation scheme given in (8) is less complex than the one given in (9) for systems with peak power constraints. However, the two schemes are equivalent in terms of outage probability, and the latter is useful for the analysis of systems with average power or PAPR constraints. In the following, we only consider $\boldsymbol{p}^{\rm peak}(\boldsymbol{\gamma})$ given in (9) for systems with peak power constraints.

The evaluation of $\boldsymbol{p}^{\rm peak}(\boldsymbol{\gamma})$ in (9) involves computing the inverse of the function $\mathrm{MMSE}_{\mathcal{X}}(\rho)$, which may be computationally prohibitive for specific practical systems. Following the analysis in [10], we obtain a suboptimal truncated water-filling power allocation rule $\boldsymbol{p}^{\rm peak}_{\rm tw}(\boldsymbol{\gamma})$ by replacing $\boldsymbol{\wp}$ in (9) with $\boldsymbol{\wp}^{\rm tw}$ given by

$$\wp_b^{\rm tw} = \min\left\{\frac{\beta}{\gamma_b}, \left(\eta - \frac{1}{\gamma_b}\right)_+\right\}, \quad b = 1, \dots, B, \tag{12}$$

where $\beta$ is a predefined design parameter[5] and $\eta$ is chosen such that the rate requirement is satisfied

$$I_B\left(\boldsymbol{p}^{\rm peak}_{\rm tw}(\boldsymbol{\gamma}), \boldsymbol{\gamma}\right) = R.$$

*2) Average Power Constraint:* Under an average power constraint, the optimal power allocation scheme solves

$$\begin{cases} \text{Minimize} & \Pr(I_B(\boldsymbol{p}(\boldsymbol{\gamma}), \boldsymbol{\gamma}) < R) \\ \text{Subject to} & \mathbb{E}\left[\langle \boldsymbol{p}(\boldsymbol{\gamma})\rangle\right] \le P_{\rm av}. \end{cases} \tag{13}$$

From [10], the solution $\boldsymbol{p}^{\rm av}(\boldsymbol{\gamma})$ of (13) is given by

$$\boldsymbol{p}^{\rm av}(\boldsymbol{\gamma}) = \begin{cases} \boldsymbol{\wp}(\boldsymbol{\gamma}), & \langle \boldsymbol{\wp}(\boldsymbol{\gamma})\rangle \le s \\ \mathbf{0}, & \text{otherwise}, \end{cases} \tag{14}$$

[5]The optimal value of $\beta$ is dependent on the transmission rate and the target outage probability. Large values of $\beta$ guarantee the optimal outage diversity over a larger range of transmission rate, while too large (and too small) values of $\beta$ decrease the achievable coding gain. See [10] for guidelines on how to find $\beta$.

where $\wp(\boldsymbol{\gamma})$ is given in (11) and $s$ is such that (noting that $\mathbb{E}\left[\langle \boldsymbol{p}^{\text{av}}(\boldsymbol{\gamma})\rangle\right]$ is a function of $s$)

$$\begin{cases} s=\infty, & \text{if } \lim_{s\to\infty}\mathbb{E}\left[\langle \boldsymbol{p}^{\text{av}}(\boldsymbol{\gamma})\rangle\right]\le P_{\text{av}} \\ P_{\text{av}}=\mathbb{E}\left[\langle \boldsymbol{p}^{\text{av}}(\boldsymbol{\gamma})\rangle\right], & \text{otherwise.} \end{cases} \tag{15}$$

The threshold $s$ is a function of $\wp(\boldsymbol{\gamma})$, $P_{\text{av}}$ and the fading statistics $f_{\boldsymbol{\gamma}}(\boldsymbol{\gamma})$; thus $s$ is fixed and can be predetermined. Consequently, the complexity of the scheme $\boldsymbol{p}^{\text{av}}(\boldsymbol{\gamma})$ is governed by the complexity of $\wp(\boldsymbol{\gamma})$. Therefore, suboptimal alternatives of $\wp(\boldsymbol{\gamma})$ can be used to reduce the complexity of $\boldsymbol{p}^{\text{av}}(\boldsymbol{\gamma})$. The truncated water-filling scheme for systems with average power constraints $\boldsymbol{p}^{\text{av}}_{\text{tw}}(\boldsymbol{\gamma})$ [10] can be obtained by employing $\wp^{\text{tw}}(\boldsymbol{\gamma})$ in stead of $\wp(\boldsymbol{\gamma})$, i.e.

$$\boldsymbol{p}^{\text{av}}_{\text{tw}}(\boldsymbol{\gamma})=\begin{cases} \wp^{\text{tw}}(\boldsymbol{\gamma}), & \langle \wp^{\text{tw}}(\boldsymbol{\gamma})\rangle\le s^{\text{tw}} \\ \mathbf{0}, & \text{otherwise,} \end{cases} \tag{16}$$

where $\wp^{\text{tw}}(\boldsymbol{\gamma})$ is given in (12) and $s^{\text{tw}}$ satisfies

$$\begin{cases} s^{\text{tw}}=\infty, & \text{if } \lim_{s^{\text{tw}}\to\infty}\mathbb{E}\left[\langle \boldsymbol{p}^{\text{av}}_{\text{tw}}(\boldsymbol{\gamma})\rangle\right]\le P_{\text{av}} \\ P_{\text{av}}=\mathbb{E}\left[\langle \boldsymbol{p}^{\text{av}}_{\text{tw}}(\boldsymbol{\gamma})\rangle\right], & \text{otherwise.} \end{cases} \tag{17}$$

### *B. Peak-to-Average Power Ratio Constraints*

For systems with average power $P_{\text{av}}$ and peak-to-average power ratio $\text{PAPR}$, the optimal power allocation scheme solves the following problem [8],

$$\begin{cases} \text{Minimize} & \Pr\left(I_B(\boldsymbol{p}(\boldsymbol{\gamma}),\boldsymbol{\gamma})<R\right) \\ \text{Subject to} & \langle \boldsymbol{p}(\boldsymbol{\gamma})\rangle\le P_{\text{peak}}=\text{PAPR}\cdot P_{\text{av}} \\ & \mathbb{E}\left[\langle \boldsymbol{p}(\boldsymbol{\gamma})\rangle\right]\le P_{\text{av}}. \end{cases} \tag{18}$$

Following the arguments in [8], the optimal power allocation rule $\boldsymbol{p}^{\star}(\boldsymbol{\gamma})$ is as follows.

*Proposition 1:* A solution to problem (18) is given by

$$\boldsymbol{p}^{\star}(\boldsymbol{\gamma}) = \begin{cases} \boldsymbol{\wp}(\boldsymbol{\gamma}), & \langle \boldsymbol{\wp}(\boldsymbol{\gamma}) \rangle \leq \hat{s} \\ \boldsymbol{0}, & \text{otherwise}, \end{cases} \tag{19}$$

where $\boldsymbol{\wp}(\boldsymbol{\gamma})$ is given in (11) and $\hat{s} = \min\{s, P_{\text{peak}}\}$ with $s$ defined as in (15).

*Proof:* If $P_{\text{av}}$ and $P_{\text{peak}}$ are such that $s \leq P_{\text{peak}}$, we have $\hat{s} = s$. Therefore, $\boldsymbol{p}^{\star}(\boldsymbol{\gamma})$ coincides with $\boldsymbol{p}^{\text{av}}(\boldsymbol{\gamma})$. Furthermore, $\boldsymbol{p}^{\star}(\boldsymbol{\gamma})$ satisfies the peak power constraint since $\langle \boldsymbol{p}^{\star}(\boldsymbol{\gamma}) \rangle \leq s \leq P_{\text{peak}}$. Consequently, $\boldsymbol{p}^{\star}(\boldsymbol{\gamma})$ is a solution of (18) since the peak power constraint is redundant.

If $P_{\text{av}}$ and $P_{\text{peak}}$ are such that $s > P_{\text{peak}}$, we have $\hat{s} = P_{\text{peak}} < s$. Therefore, $\boldsymbol{p}^{\star}(\boldsymbol{\gamma})$ coincides with $\boldsymbol{p}^{\text{peak}}(\boldsymbol{\gamma})$. Now, noting that $\mathbb{E}\left[\langle \boldsymbol{p}^{\star}(\boldsymbol{\gamma}) \rangle\right]$ is an increasing function of $\hat{s}$, we have that $\mathbb{E}\left[\langle \boldsymbol{p}^{\star}(\boldsymbol{\gamma}) \rangle\right] < \mathbb{E}\left[\langle \boldsymbol{p}^{\text{av}}(\boldsymbol{\gamma}) \rangle\right] \leq P_{\text{av}}$. Consequently, $\boldsymbol{p}^{\star}(\boldsymbol{\gamma})$ is a solution of (18) since the average power constraint is redundant.

Thus, in all cases, $\boldsymbol{p}^{\star}(\boldsymbol{\gamma})$ is a solution of (18). ■

*Remark 1:* From the proof, we observe that, depending on $P_{\text{av}}$ and the PAPR (which is fixed), one of the power constraints is redundant and the outage performance is dependent on the remaining constraint. In particular we have that

$$P_{\text{out}}\left(\boldsymbol{p}^{\star}(\boldsymbol{\gamma}), P_{\text{av}}, R\right) = \begin{cases} P_{\text{out}}\left(\boldsymbol{p}^{\text{peak}}(\boldsymbol{\gamma}), P_{\text{peak}}, R\right), & s > P_{\text{peak}} \\ P_{\text{out}}\left(\boldsymbol{p}^{\text{av}}(\boldsymbol{\gamma}), P_{\text{av}}, R\right), & s \leq P_{\text{peak}}. \end{cases} \tag{20}$$

Consequently, the outage probability can also be evaluated as

$$\begin{aligned} P_{\text{out}}\left(\boldsymbol{p}^{\star}(\boldsymbol{\gamma}), P_{\text{av}}, R\right) &= \max\left\{P_{\text{out}}\left(\boldsymbol{p}^{\text{peak}}(\boldsymbol{\gamma}), P_{\text{peak}}, R\right), P_{\text{out}}\left(\boldsymbol{p}^{\text{av}}(\boldsymbol{\gamma}), P_{\text{av}}, R\right)\right\} \\ &= \max\left\{P_{\text{out}}\left(\boldsymbol{p}^{\text{peak}}(\boldsymbol{\gamma}), \text{PAPR} \cdot P_{\text{av}}, R\right), P_{\text{out}}\left(\boldsymbol{p}^{\text{av}}(\boldsymbol{\gamma}), P_{\text{av}}, R\right)\right\}. \end{aligned} \tag{21}$$

The above expression clearly highlights that in order to compute the outage probability with PAPR constraints, it is sufficient to translate the curve corresponding to the peak power constraint by $\text{PAPR}$ dB and then find the maximum between the translated curve and the curve corresponding to the average power constraint.

With similar arguments to the previous section, the suboptimal truncated water-filling scheme $\boldsymbol{p}_{\rm tw}^{\star}(\boldsymbol{\gamma})$ for systems with PAPR constraints is given by

$$\boldsymbol{p}_{\rm tw}^{\star}(\boldsymbol{\gamma}) = \begin{cases} \wp^{\rm tw}(\boldsymbol{\gamma}), & \langle \wp^{\rm tw}(\boldsymbol{\gamma}) \rangle \le \hat{s}^{\rm tw} \\ \mathbf{0}, & \text{otherwise}, \end{cases} \tag{22}$$

with $\hat{s}^{\rm tw} = \min\{s^{\rm tw}, P_{\rm peak}\}$ where $s^{\rm tw}$ is given in (17). The outage probability of systems with PAPR constraints is also given by

$$P_{\rm out}(\boldsymbol{p}_{\rm tw}^{\star}(\boldsymbol{\gamma}), P_{\rm av}, R) = \max\Big\{P_{\rm out}\left(\boldsymbol{p}_{\rm tw}^{\rm peak}(\boldsymbol{\gamma}), {\rm PAPR} \cdot P_{\rm av}, R\right), P_{\rm out}\left(\boldsymbol{p}_{\rm tw}^{\rm av}(\boldsymbol{\gamma}), P_{\rm av}, R\right)\Big\}.$$

*1) Asymptotic Analysis:* In this section we study the asymptotic behavior of the outage probability under PAPR constraints. In particular, we study the SNR exponents, i.e., the asymptotic slope of the outage probability for large SNR. For large $P_{\rm av}$, we have the following result.

*Proposition 2:* Consider transmission at rate $R$ over the block-fading channel given in (1) with power allocation scheme $\boldsymbol{p}^{\star}(\boldsymbol{\gamma})$ (or $\boldsymbol{p}_{\rm tw}^{\star}(\boldsymbol{\gamma})$). Assume input constellation $\mathcal{X}$ of size $2^M$. Further assume that the power fading gains $\boldsymbol{\gamma}$ follow the Nakagami-$m$ distribution given in (2). Then, for large $P_{\rm av}$ and any ${\rm PAPR} < \infty$, the outage probability behaves like

$$P_{\rm out}\left(\boldsymbol{p}^{\star}(\boldsymbol{\gamma}), P_{\rm av}, R\right) \doteq \mathcal{K} P_{\rm av}^{-md(R)} \tag{23}$$

$$P_{\rm out}\left(\boldsymbol{p}_{\rm tw}^{\star}(\boldsymbol{\gamma}), P_{\rm av}, R\right) \doteq \mathcal{K}_{\beta} P_{\rm av}^{-md_{\beta}(R)}, \tag{24}$$

where $d(R)$ is the Singleton bound [19], [20], [21], [22]

$$d(R) = 1 + \left\lfloor B\left(1 - \frac{R}{M}\right)\right\rfloor,$$

and $d_{\beta}(R)$ is given by

$$d_{\beta}(R) = 1 + \left\lfloor B\left(1 - \frac{R}{I_{\mathcal{X}}(\beta)}\right)\right\rfloor. \tag{25}$$

*Proof:* For sufficiently large $P_{\rm peak}$ we have that [10]

$$P_{\rm out}\left(\boldsymbol{p}^{\rm peak}(\boldsymbol{\gamma}), P_{\rm peak}, R\right) \doteq \mathcal{K}^{\rm peak} P_{\rm peak}^{-md(R)}. \tag{26}$$

Let $P(s)$ be the average power constraint as a function of the threshold $s$ in the allocation scheme $\boldsymbol{p}^{\text{av}}(\boldsymbol{\gamma})$ in (14). Asymptotically with $s$, we have [10]

$$\frac{d}{ds}P(s) \doteq \mathcal{K}^{\text{peak}} d(R) s^{-d(R)}.$$

From L'Hôpital's rule, we have for any $\text{PAPR}$

$$\lim_{s\to\infty} \frac{\text{PAPR}\cdot P(s)}{s} = \lim_{s\to\infty} \frac{d}{ds}\text{PAPR}\cdot P(s) = \lim_{s\to\infty} \text{PAPR}\cdot \mathcal{K} d(R) s^{-d(R)} = 0.$$

It follows that for any $\text{PAPR}$, there exists an $s_0$ and the corresponding average power constraint $P_0 = P(s_0)$ such that $s_0 = \text{PAPR}\cdot P_0$ and $s > P(s)\cdot\text{PAPR}$ if $P(s) > P_0$. Consequently, $P_{\text{out}}\left(\boldsymbol{p}^{\star}(\boldsymbol{\gamma}), P_{\text{av}}, R\right) = P_{\text{out}}\left(\boldsymbol{p}^{\text{peak}}(\boldsymbol{\gamma}), \text{PAPR}\cdot P_{\text{av}}, R\right)$ for $P_{\text{av}} > P_0$. Thus, together with (26), at large $P_{\text{av}}$, we have

$$P_{\text{out}}\left(\boldsymbol{p}^{\star}(\boldsymbol{\gamma}), P_{\text{av}}, R\right) \doteq P_{\text{out}}\left(\boldsymbol{p}^{\text{peak}}(\boldsymbol{\gamma}), \text{PAPR}\cdot P_{\text{av}}, R\right) \doteq \mathcal{K}^{\text{peak}}\text{PAPR}^{-md(R)} P_{\text{av}}^{-md(R)} \tag{27}$$

as stated in (23).

By noting that [10]

$$P_{\text{out}}\left(\boldsymbol{p}_{\text{tw}}^{\text{peak}}(\boldsymbol{\gamma}), P_{\text{peak}}, R\right) \doteq \mathcal{K}_{\beta}^{\text{peak}} P_{\text{peak}}^{-md_{\beta}(R)},$$

the proof for the suboptimal scheme $\boldsymbol{p}_{\text{tw}}^{\star}(\boldsymbol{\gamma})$ follows using the same arguments as above. ∎

The threshold $P_0$ in the proof is the average power constraint such that the threshold $s$ in (15) satisfies $s = \text{PAPR}\cdot P_0$. Equivalently, $P_0$ satisfies

$$\int_{\boldsymbol{\gamma}:\langle \boldsymbol{\wp}(\boldsymbol{\gamma})\rangle \le \text{PAPR}\cdot P_0} \langle \boldsymbol{\wp}(\boldsymbol{\gamma})\rangle \, dF_{\boldsymbol{\gamma}}(\boldsymbol{\gamma}) = P_0, \tag{28}$$

where $F_{\boldsymbol{\gamma}}(\boldsymbol{\gamma})$ is the joint pdf of $\boldsymbol{\gamma} = (\gamma_1, \ldots, \gamma_B)$. We therefore have that

$$P_{\text{out}}\left(\boldsymbol{p}^{\star}(\boldsymbol{\gamma}), P_{\text{av}}, R\right) = P_{\text{out}}\left(\boldsymbol{p}^{\text{peak}}(\boldsymbol{\gamma}), \text{PAPR}\cdot P_{\text{av}}, R\right)$$

for $P_{\text{av}} > P_0$. Therefore, for asymptotically large $P_{\text{av}}$, the outage probability for systems with a PAPR constraint is determined by the outage probability of systems with peak power constraint $P_{\text{peak}} = \text{PAPR}\cdot P_{\text{av}}$. As a consequence of the above analysis, we have that the delay-limited capacity [23] is zero for any finite $\text{PAPR}$. This is illustrated by examples in the next section.

*2) Numerical Results:* For simplicity, we first consider the outage performance of systems with $B = 1$ under Nakagami-$m$ fading statistic. Then, the outage probability can be numerically evaluated as follows. Let $\gamma$ be the power fading gain, then $\wp(\gamma) = \frac{I_{\mathcal{X}}^{-1}(R)}{\gamma}$ and (28) reduces to

$$\begin{aligned}
\int_{\frac{I_{\mathcal{X}}^{-1}(R)}{P_0 \cdot \mathrm{PAPR}}}^{\infty} \frac{I_{\mathcal{X}}^{-1}(R)}{\gamma} \frac{m^m \gamma^{m-1}}{\Gamma(m)} e^{-m\gamma} d\gamma &= P_0 \\
\frac{m^m}{\Gamma(m)} a \int_{\frac{a}{\mathrm{PAPR}}}^{\infty} \gamma^{m-2} e^{-m\gamma} d\gamma &= 1 \\
\frac{m}{\Gamma(m)} a \Gamma\left(m-1, m\frac{a}{\mathrm{PAPR}}\right) &= 1, \qquad (29)
\end{aligned}$$

where $a \triangleq \frac{I_{\mathcal{X}}^{-1}(R)}{P_0}$ and $\Gamma(n,\xi)$ is the upper incomplete Gamma function [24] defined as $\Gamma(n,\xi) \triangleq \int_{\xi}^{\infty} t^{n-1} e^{-t} dt$.

The threshold $P_0$ can be obtained by solving (29) for $a$. For $P_{\mathrm{av}} > P_0$ $(s > \mathrm{PAPR} \cdot P_{\mathrm{av}})$ the outage probability is given by

$$P_{\mathrm{out}}\left(\boldsymbol{p}^{\mathrm{peak}}(\boldsymbol{\gamma}), \mathrm{PAPR} \cdot P_{\mathrm{av}}, R\right) = \Pr\left(\gamma < \frac{I_{\mathcal{X}}^{-1}(R)}{\mathrm{PAPR} \cdot P_{\mathrm{av}}}\right) = F_{\gamma}\left(\frac{I_{\mathcal{X}}^{-1}(R)}{\mathrm{PAPR} \cdot P_{\mathrm{av}}}\right).$$

For $P_{\mathrm{av}} < P_0$ $(s < \mathrm{PAPR} \cdot P_{\mathrm{av}})$, $s$ in (14) is obtained by solving

$$\frac{m I_{\mathcal{X}}^{-1}(R)}{\Gamma(m)} \Gamma\left(m-1, \frac{m I_{\mathcal{X}}^{-1}(R)}{s}\right) = P_{\mathrm{av}},$$

and the outage probability is given by

$$P_{\mathrm{out}}\left(\boldsymbol{p}^{\mathrm{av}}(\boldsymbol{\gamma}), P_{\mathrm{av}}, R\right) = \Pr\left(\gamma < \frac{I_{\mathcal{X}}^{-1}(R)}{s}\right) = F_{\gamma}\left(\frac{I_{\mathcal{X}}^{-1}(R)}{s}\right).$$

The analytical result for $B = 1$ is illustrated in Figure 1 for a 16-QAM input, Rayleigh fading channel at rate $R = 1$. We observe that as we increase the PAPR constraint, the error floor occurs at lower error probability values, and eventually, at values below a target quality-of-service error rate. We also observe that the loss incurred by BICM is minimal.

For systems with $B > 1$, analytical results are not available in closed form. However, from (21), the outage probability of systems with PAPR constraints can be obtained by considering systems with peak power constraints and systems with average power constraints separately. Moreover, at high $P_{\text{av}}$, the outage probability can be obtained by the outage probability of systems with only a peak power constraint $P_{\text{av}} \cdot \text{PAPR}$. Simulation results for a 16-QAM input, Rayleigh fading channel with $B = 4$ blocks at rate $R = 3$ are given in Figure 2.

In both cases ($B = 1$ and $B = 4$), the outage probability at high $P_{\text{av}}$ resulting from the optimal power allocation scheme is governed by the peak power constraints, and therefore, the optimal outage diversity is given by the Singleton bound.

The outage performance of systems with 16-QAM inputs, Rayleigh fading channel with $B = 4, R = 3$, employing the truncated water-filling scheme is illustrated in Figure 3. It follows from (24) that $\beta = \infty$ is required to maintain the optimal diversity. Therefore, we need to choose $\beta$ relatively high ($\beta = 19$dB) to keep the outage performance close to optimal at outage probability $10^{-5}$. The suboptimal outage diversity $d_\beta(R) = d(R) - 1$ appears at lower outage probability. For rates $R$ such that $B\left(1 - \frac{R}{M}\right)$ is not an integer, optimal diversity can be maintained with finite $\beta$, thus smaller values of $\beta$ can be chosen, which results in smaller performance gap between the truncated water-filling and the optimal scheme.

## V. Ergodic Capacity Maximization

We now consider the capacity of the ergodic channel, where the number of block $B$ is sufficiently large to reveal the statistics of the channel within one codeword. The channel model follows from (1) by letting $B \to \infty$ and $L = 1$. For a given power allocation rule $p(\gamma)$, the ergodic capacity of the channel is

$$C = \mathbb{E}_\gamma \left[ I_{\mathcal{X}}(p(\gamma)\gamma) \right] = \int_{\gamma > 0} I_{\mathcal{X}}(p(\gamma)\gamma) f_\gamma(\gamma) d\gamma. \tag{30}$$

Similarly to the previous section, we first preview the channel capacity under average power constraints before presenting the results on the channel capacity under PAPR constraints.

### A. Average Power Constraint

For a system with an average power constraint $P_{\rm av}$, the optimal power allocation rule is given by

$$p^{\rm opt}(\gamma) = \arg \max_{\mathbb{E}_\gamma[p(\gamma)] \leq P_{\rm av}} \mathbb{E}_\gamma \left[ I_{\mathcal{X}}(p(\gamma)\gamma) \right]. \tag{31}$$

The solution is given by [9]

$$p^{\rm opt}(\gamma) = \frac{1}{\gamma} {\rm MMSE}_{\mathcal{X}}^{-1} \left( \min \left\{ {\rm MMSE}_{\mathcal{X}}(0), \frac{\eta}{\gamma} \right\} \right), \tag{32}$$

where $\eta$ is chosen such that $\mathbb{E}_\gamma \left[ p^{\rm opt}(\gamma) \right] = P_{\rm av}$. The resulting capacity is

$$C^{\rm opt} = \int_{\frac{\eta}{{\rm MMSE}_{\mathcal{X}}(0)}}^{\infty} I_{\mathcal{X}} \left( {\rm MMSE}_{\mathcal{X}}^{-1} \left( \frac{\eta}{\gamma} \right) \right) f_\gamma(\gamma) d\gamma. \tag{33}$$

A low-complexity suboptimal solution to problem (31) of $p^{\rm opt}(\gamma)$ can be derived by approximating $I_{\mathcal{X}}(\rho)$ with the following bound

$$I_{\mathcal{X}}^{u}(\rho) = \min\{\log_2(1+\rho), \log_2(1+\beta)\}, \tag{34}$$

where $\beta$ is a predefined parameter to be optimized depending on $P_{\rm av}$. The suboptimal power allocation scheme is given by

$$p^{\rm tw}(\gamma) = \arg \max_{\mathbb{E}_\gamma[p(\gamma)] \leq P_{\rm av}} \mathbb{E}_\gamma \left[ I_{\mathcal{X}}^{u}(p(\gamma)\gamma) \right]. \tag{35}$$

Since $I_{\mathcal{X}}^{u}(p(\gamma)\gamma) = \log_2(1+\beta)$ if $p(\gamma) \geq \frac{\beta}{\gamma}$, the solution of (35) satisfies $p(\gamma) \leq \frac{\beta}{\gamma}$. Therefore, (35) is equivalent to

$$p^{\rm tw}(\gamma) = \arg \max_{\substack{\mathbb{E}_\gamma[p(\gamma)] \leq P_{\rm av} \\ p(\gamma) \leq \frac{\beta}{\gamma}}} \mathbb{E}_\gamma \left[ \log_2(1+p(\gamma)\gamma) \right]. \tag{36}$$

Using the Karush-Kurhn-Tucker (KKT) conditions, we have that

$$p^{\rm tw}(\gamma) = \min \left\{ \frac{\beta}{\gamma}, \left( \eta - \frac{1}{\gamma} \right)_+ \right\}, \tag{37}$$

where $\eta$ is chosen such that $\mathbb{E}_\gamma \left[ p^{\rm tw}(\gamma) \right] = P_{\rm av}$. The resulting capacity is

$$C^{\rm tw} = \int_{\frac{1}{\eta}}^{\frac{\beta+1}{\eta}} I_{\mathcal{X}}(\eta\gamma - 1) f_\gamma(\gamma) d\gamma + I_{\mathcal{X}}(\beta) \left( 1 - F_\gamma \left( \frac{\beta+1}{\eta} \right) \right). \tag{38}$$

### B. Peak-to-Average Power Constraint

For systems with a PAPR constraints, the optimal power allocation rule is given by

$$p_{\text{papr}}^{\text{opt}}(\gamma) = \arg \max_{\substack{\mathbb{E}_\gamma[p(\gamma)] \le P_{\text{av}} \\ p(\gamma) \le P_{\text{peak}}}} \mathbb{E}_\gamma \left[ I_{\mathcal{X}}(p(\gamma)\gamma) \right], \tag{39}$$

where $P_{\text{peak}} = \text{PAPR} \cdot P_{\text{av}}$. Applying the KKT conditions, the optimal power allocation scheme is given by

$$p_{\text{papr}}^{\text{opt}}(\gamma) = \min \left\{ P_{\text{peak}}, \frac{1}{\gamma} \text{MMSE}_{\mathcal{X}}^{-1} \left( \min \left\{ \text{MMSE}_{\mathcal{X}}(0), \frac{\eta}{\gamma} \right\} \right) \right\}, \tag{40}$$

where $\eta$ is chosen such that $\mathbb{E}_\gamma \left[ p_{\text{papr}}^{\text{opt}}(\gamma) \right] = P_{\text{av}}$.

Similarly to the previous section, we derive a suboptimal power allocation rule based on the truncated water-filling algorithm by solving

$$p_{\text{papr}}^{\text{tw}}(\gamma) = \arg \max_{\substack{\mathbb{E}_\gamma[p(\gamma)] \le P_{\text{av}} \\ p(\gamma) \le \min\left\{ P_{\text{peak}}, \frac{\beta}{\gamma} \right\}}} \mathbb{E}_\gamma \left[ \log_2(1 + p(\gamma)\gamma) \right]. \tag{41}$$

Let $\alpha(\gamma) = \min \left\{ P_{\text{peak}}, \frac{\beta}{\gamma} \right\}$, then a truncated water filling suboptimal of $p_{\text{papr}}^{\text{opt}}$ is given by

$$p_{\text{papr}}^{\text{tw}}(\gamma) = \min \left\{ \alpha(\gamma), \left( \eta - \frac{1}{\gamma} \right)_+ \right\}, \tag{42}$$

where $\eta$ is chosen such that $\mathbb{E}_\gamma \left[ p_{\text{papr}}^{\text{tw}}(\gamma) \right] = P_{\text{av}}$. It can be seen that if $\eta \le P_{\text{peak}}$ or $\frac{\beta+1}{\eta} \le \frac{1}{\eta - P_{\text{peak}}}$, (42) is equivalent to (37). Therefore, the resulting ergodic capacity is given in (33). Otherwise, let $a = \frac{1}{\eta - P_{\text{peak}}}$ and $b = \frac{\beta}{P_{\text{peak}}}$, then the resulting ergodic capacity can be written as

$$C_{\text{papr}}^{\text{tw}} = \int_{1/\eta}^{a} I_{\mathcal{X}}(\eta\gamma - 1) f_\gamma(\gamma) d\gamma + \int_{a}^{b} I_{\mathcal{X}}(P_{\text{peak}}\gamma) f_\gamma(\gamma) d\gamma + \left(1 - F_\gamma(b)\right) I_{\mathcal{X}}(\beta). \tag{43}$$

### C. Numerical Results

The capacities presented in the previous sections can easily be calculated using Gaussian quadrature integrations. Numerical results for the ergodic capacity of Rayleigh fading channels with 16-QAM inputs are presented in Figures 4, 5, 6. Figures 4 and 5 show the performance of the truncated water-filling scheme with average power constraints and PAPR constraints, respectively, where $\beta$ has been chosen to maximize capacity at each $P_{\rm av}$. The results show that the truncated water-filling scheme are very close to optimal for both systems with average power constraints and systems with PAPR constraints. The truncated water-filling scheme is therefore a potential candidate for practical system implementation due to the very low computational and storage requirements compared to the optimal scheme. Figure 6 shows the ergodic capacity for various PAPR constraints. We observe that minimal loss in capacity is incurred, even with relatively small PAPR.

## VI. Conclusions

We have studied power allocation schemes under PAPR constraints for ergodic and delay-limited block-fading channels with arbitrary input distributions. In each case, we have computed the optimal solution and proposed a suboptimal scheme that requires lower computational and storage capabilities while performing close to optimal. In the delay-limited block-fading case, we have shown that the optimal and suboptimal solutions can be easily computed from the corresponding solutions with independent peak and average power constraints. We have studied the SNR exponents, and shown that the asymptotic performance for finite PAPR is always determined by the peak power, and the exponent is therefore given by the exponent of systems with peak power constraints. In the ergodic case, we have seen that even small PAPR values entail minimal capacity loss.

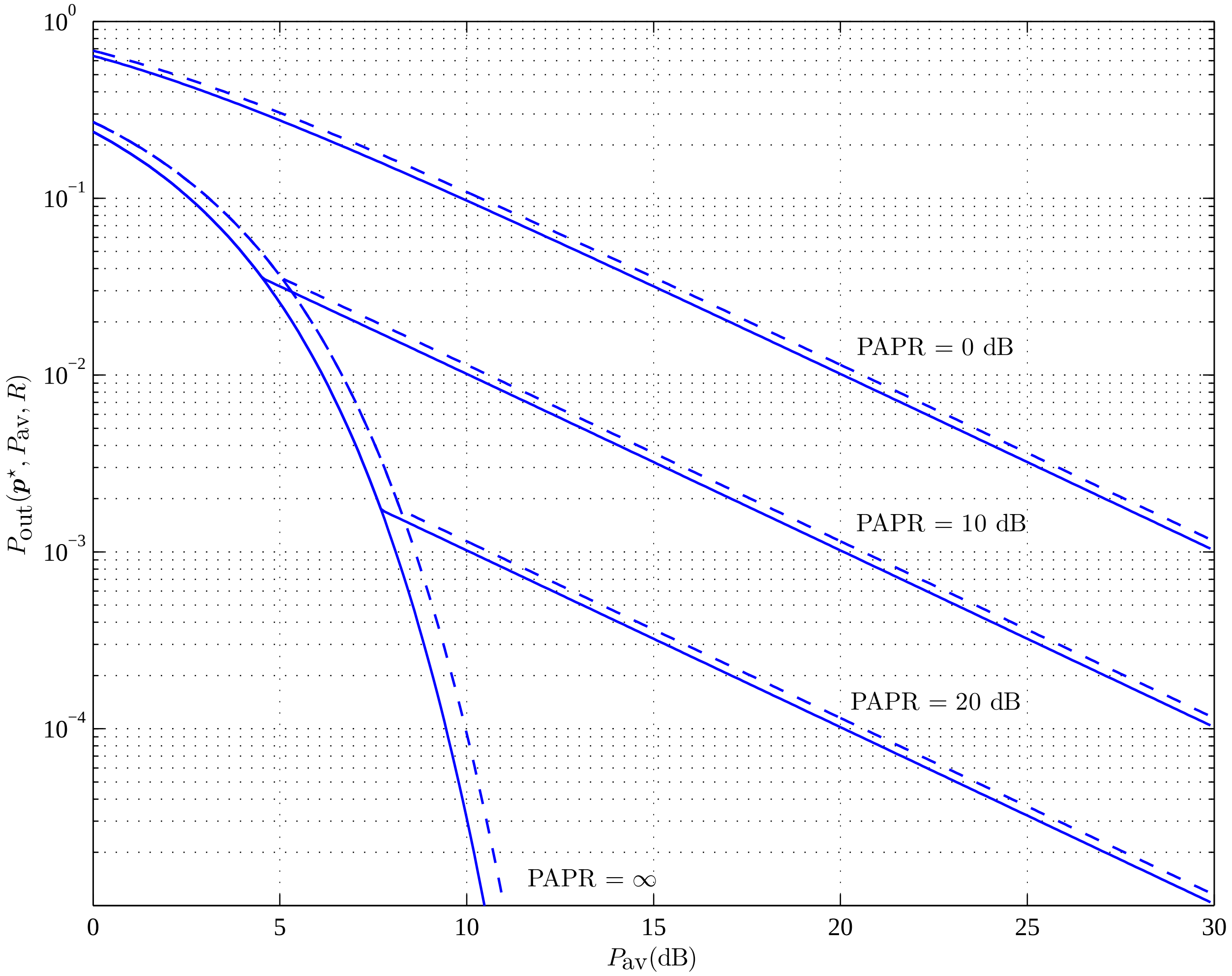

Fig. 1. Outage probability for systems with PAPR constraints over Nakagami-$m$ block-fading channels $B = 1, m = 1, R = 1$, 16-QAM inputs. The solid and dashed lines correspondingly represent outage probability of systems with coded modulation and BICM.

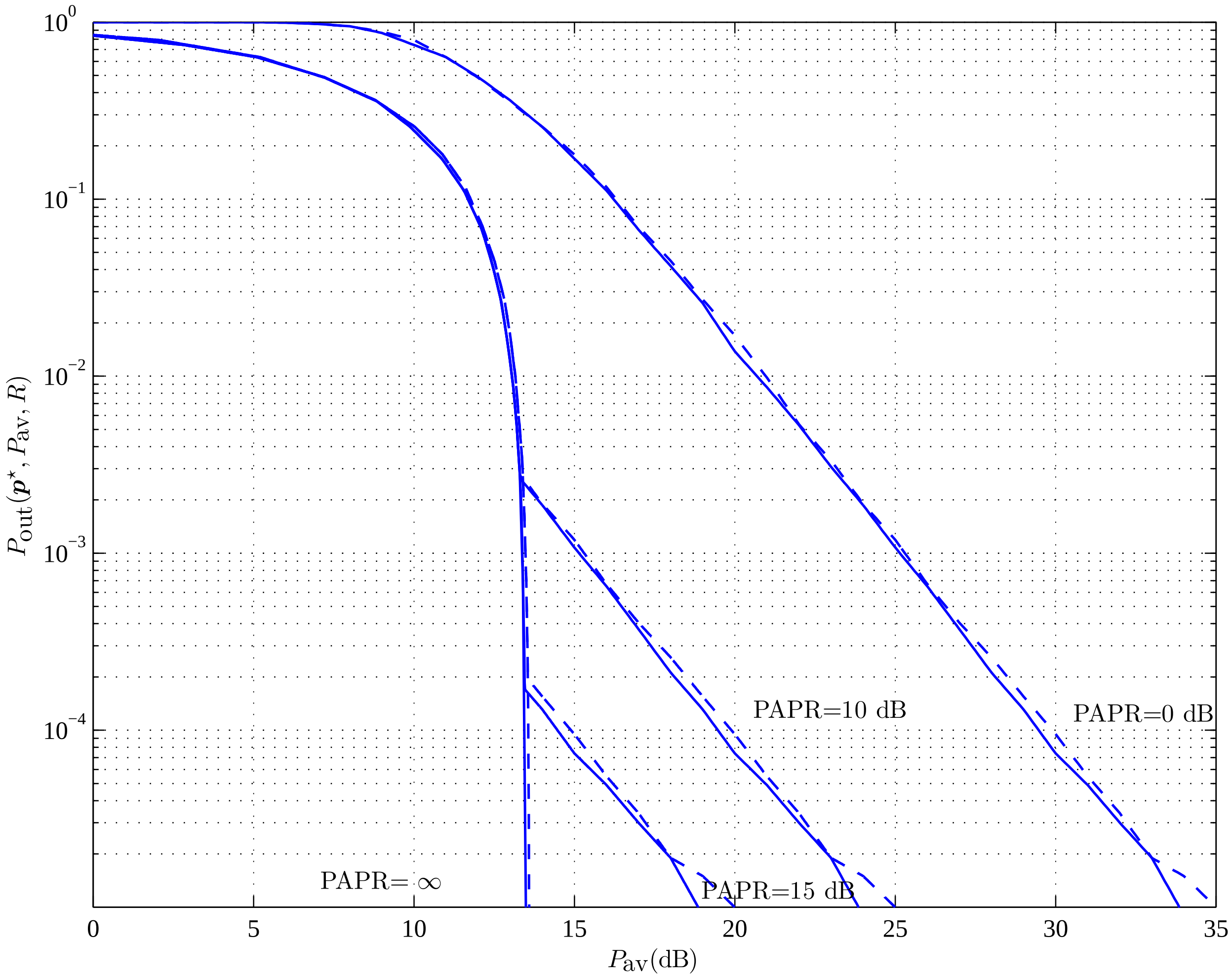

Fig. 2. Outage probability for systems with PAPR constraints over Nakagami-$m$ block-fading channels $B = 4, m = 1, R = 3$, 16-QAM inputs. The solid and dashed lines correspondingly represent outage probability of systems with coded modulation and BICM.

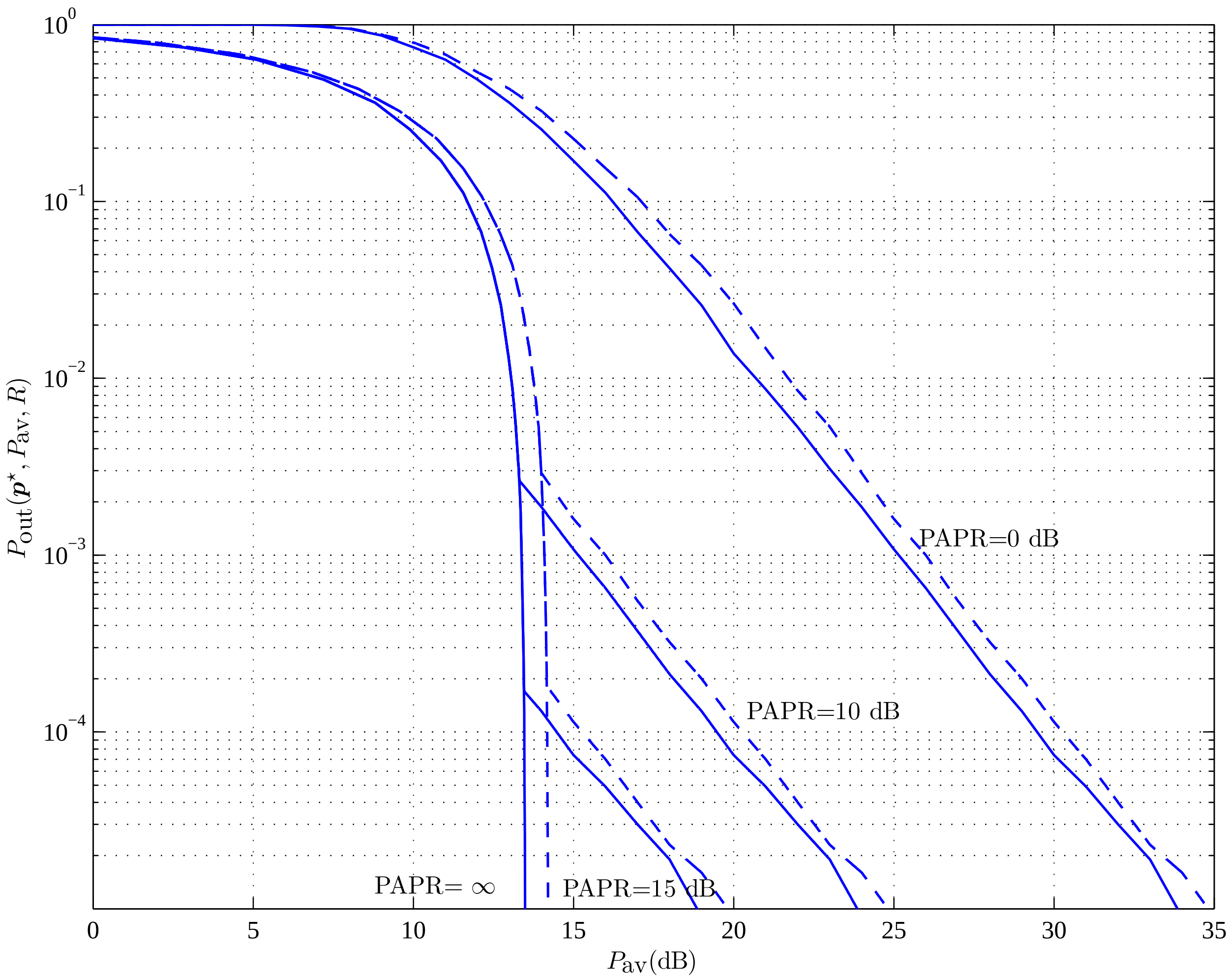

Fig. 3. Outage probability for systems with peak and average power constraints using 16-QAM input constellation over Nakgami-$m$ block-fading channels with $B = 4, m = 1, R = 3$ and peak-to-average power ratio PAPR. The solid and dashed lines correspondingly represent outage probability of systems with optimal and truncated water-filling schemes with $\beta = 19$ dB.

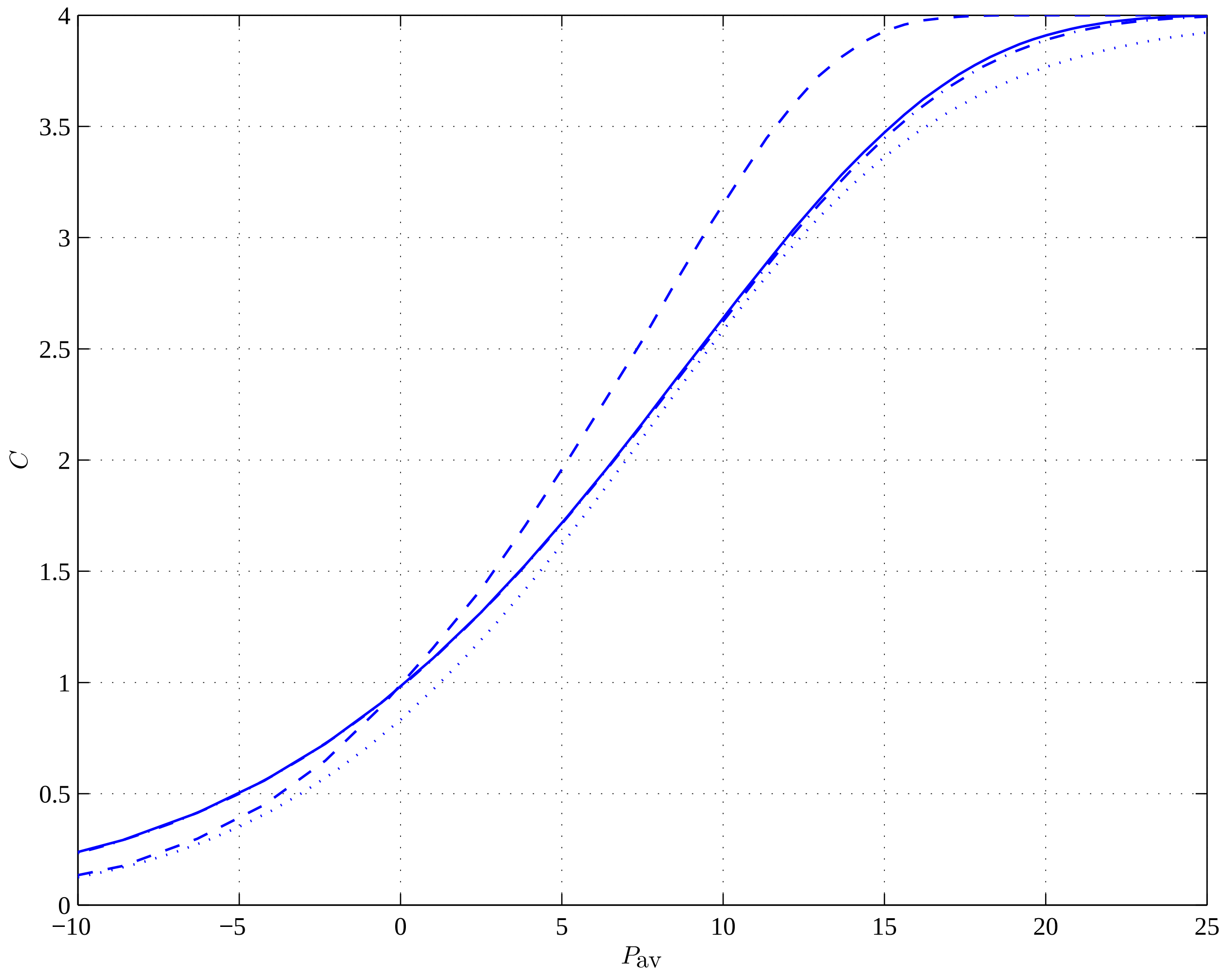

Fig. 4. Capacity of ergodic fading channel with $m = 1$, 16-QAM coded modulation inputs and average power constraint. The dashed line represents capacity of the unfaded AWGN channel, and the solid, dashed-dotted and dotted lines correspondingly represent capacities with optimal, truncated water-filling and uniform power allocation.

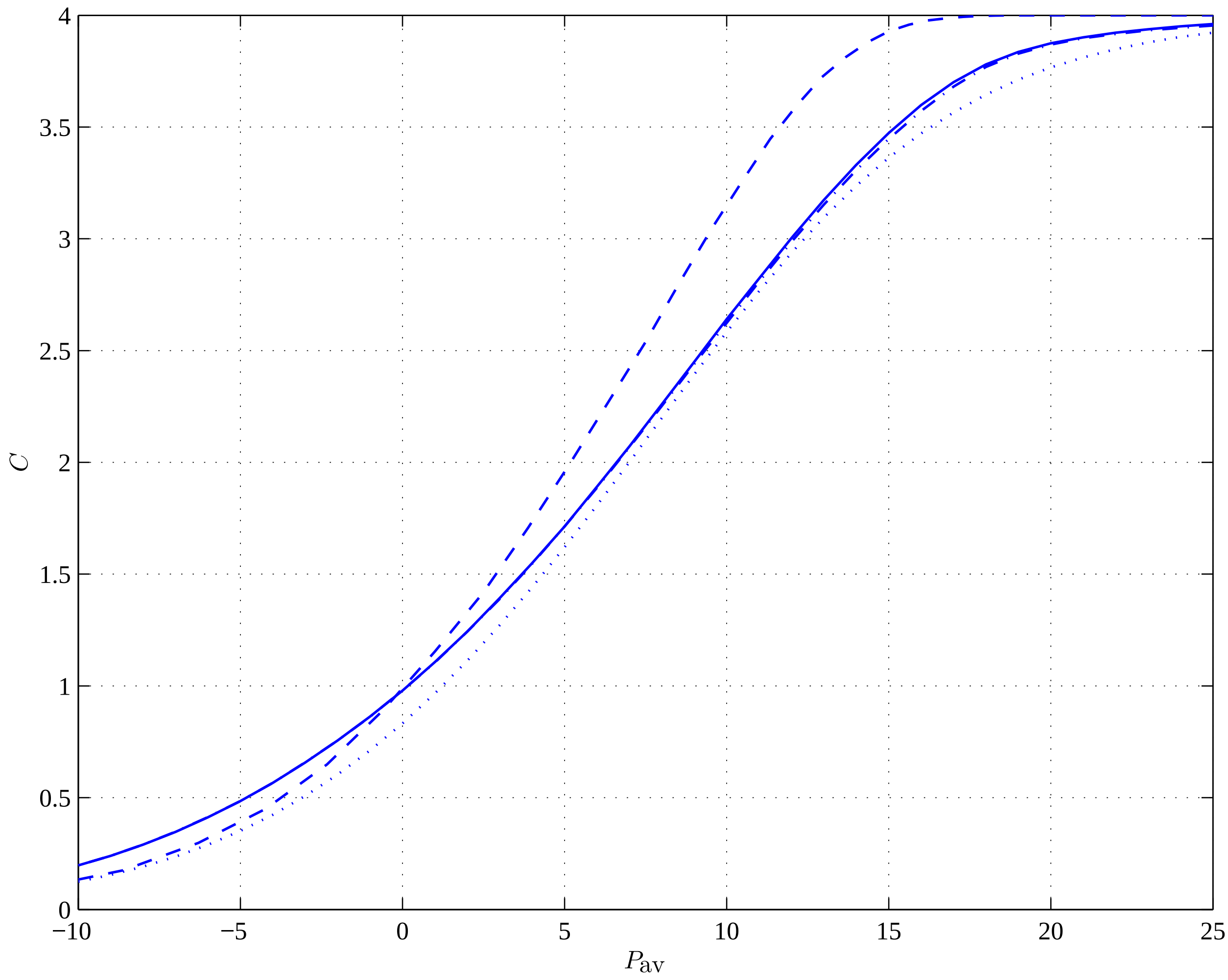

Fig. 5. Capacity of ergodic fading channel with $m = 1$, 16-QAM coded modulation inputs and $\mathrm{PAPR} = 3$dB. The dashed line represents capacity of the unfaded AWGN channel and the solid, dashed-dotted and dotted lines correspondingly represent capacities with optimal, truncated water-filling and uniform power allocation.

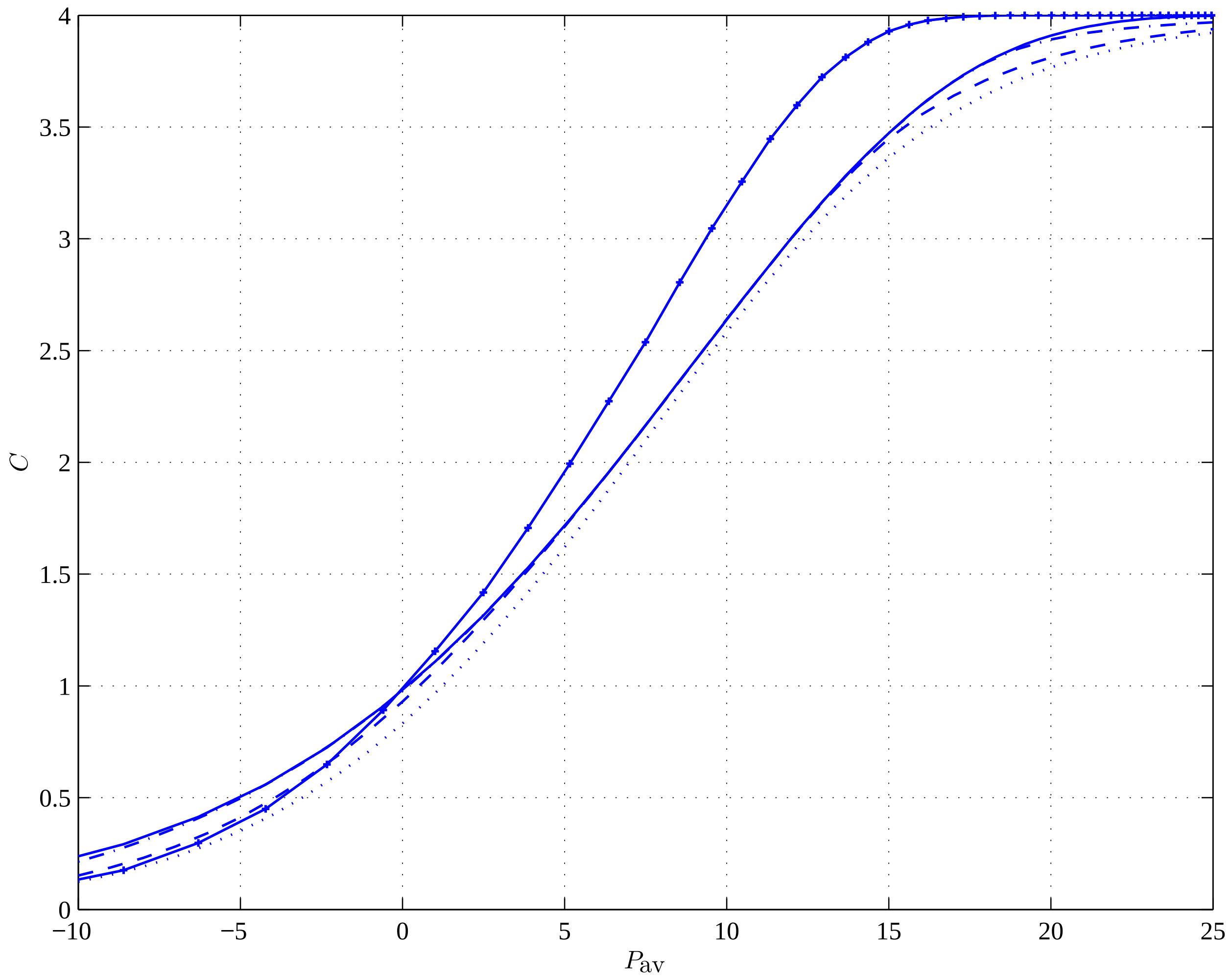

Fig. 6. Capacity of ergodic fading channel with $m = 1$, 16-QAM coded modulation inputs and PAPR constraints. The solid line with crosses represents capacity of the unfaded AWGN channel; the dotted line represents the capacity with uniform power allocation and the solid, dashed-dotted and dashed lines correspondingly represent capacities with $\mathrm{PAPR} = \infty, 4, 1$dB.